\documentclass[12pt]{article}
\usepackage{epsfig}
\newcommand{\be}{\begin{equation}}
\newcommand{\ee}{\end{equation}}
\newcommand{\bea}{\begin{eqnarray}}
\newcommand{\eea}{\end{eqnarray}}

\begin{document}

\baselineskip=.33in

\begin{flushright}
LA-UR-03-2056 \\
compiled \today
\end{flushright}

\begin{center}

\vspace{1cm}

{\large{\bf Resource Note on Photofission of Nuclei \\
for $\mathbf{^{235}}$U and $\mathbf{^{239}}$Pu Detection   }}\\[7mm]

Michael Martin Nieto \\
{\it Theoretical Division (T-8, MS-B285), 
Los Alamos National Laboratory,\\
University of California, 
Los Alamos, New Mexico 87545, U.S.A.}\\[5mm]

\end{center}

\vspace{1in}





\begin{center}
{\bf ABSTRACT}	
\end{center}

Open-source data exists, in widely scattered places, on photofission
of the important nuclear isotopes $^{235}$U and $^{239}$Pu.  
This data is useful for studies aimed at detecting
these materials at ports of entry. An 
introductory survey is given to access that data.

\newpage

\section{Introduction}

Mindful that it is now necessary to protect against the surreptitious
importation of the nuclear materials  $^{235}$U and $^{239}$Pu, 
various schemes are being proposed to detect them at borders. 
An example is that of Little et al. \cite{1-1}. They
are studying the use of a beam of photons, with energy of say $\le 10$
MeV, to survey incoming freight and use the delayed neutrons from
photofission of these nuclei as a signature of suspicious materials.

Therefore, existing data on (i) the photon cross-sections on these
nuclei, (ii) photofission products, and (iii) the delayed neutron
energy spectra are all of great interest.  Although there exist 
classical, older surveys of data \cite{1-2} - \cite{1-4}, it is
important to collate more modern results.  


\section{Photon cross-sections on $^{235}$U and  $^{239}$Pu}

Modern experiments to measure the photofission cross sections for
$^{235}$U and $^{239}$Pu began to produce results in 1980.  The first
success was by Berman's collaboration, using  
$^{235}$U.  The total cross section and the
separate fission cross sections for 1, 2, 3 neutrons were obtain for
photon energies up to 20 MeV \cite{2-1}.  (See especially Figure 2 of
\cite{2-1}.)  Further work was also done on the photoneutron
multiplicities from monoenergetic photons from 5.5 to 18 MeV 
\cite{2-2}.  

A later effort by the same collaboration produced similar results for 
$^{239}$Pu \cite{2-3}.  For example, in Figure 7 of \cite{2-3} 
the total cross section and the
separate fission cross sections for 1, 2, 3 neutrons for
photon energies up to 20 MeV are given.  

These and other results are compiled in the modern version of
Ref. \cite{1-4}, the  ``Atlas of Photoneutron Cross Sections
Obtained with Monoenergetic Photons'' \cite{2-4}. 


\section{Photofission products}

\hspace{.25in} $\mathbf{^{235}U:}$  
Starting in 1976 \cite{b-0}, a great deal of work on the
photofission products of $^{235}$U was done by the Belgium group in
Gent,  mainly in the 12-30 MeV photon energy range.  
There were detailed studies on the fragment mass distributions 
\cite{b-1}, the charge distributions \cite{b-2}, and on the isotopic and
elemental yields \cite{b-3}.  As an example of the large amount of
information in these papers, note that in table II of \cite{b-3}, the
percentage elemental yields of Kr to Ha from $^{235}$U are given.  
Isotopic information is also available.  

More specific studies on the isomeric yield ratios for  different
elements were also done \cite{b-4,b-5}.  

$\mathbf{^{239}Pu:}$  Kinetic energy and fragment mass
distributions from photofission on $^{239}$Pu were also studied by the
Gent group \cite{b-6}.   But in addition, almost simultaneously a
study by a Russian group also appeared \cite{nb-7}.  This gave 
element product
yields from a bremsstrahlung beam of maximum energy up to 28 MeV.  Both
sources should be consulted.   

{\bf Both} $\mathbf{^{235}U}$ {\bf and} $\mathbf{^{239}Pu:}$
Later, there were two non-western studies on both nuclei.  The first,
from a Japanese group \cite{nb-8}, was interested in the transmutation of
high-level radioactive waste.  It used 20, 30 and 60 MeV
bremsstrahlung. They found results within 10\% agreement with
calculated values using published photonuclear cross-section data. 
Another study, by a Russian group from Obminsk, used 
photon energies up to 11 MeV.  Their study found the yields for a number of
odd nuclei, including both $^{235}$U and $^{239}$Pu \cite{nb-9}.  
These two studies and 
their included references are a valuable resource to an audience that
might not be acquainted with this literature.   
 

\section{Delayed neutron energy spectra}

The final category of critical information deals with the delayed
neutron spectra from photofissioned $^{235}$U and $^{239}$Pu.  
The literature is replete with many studies on this problem.  See,
e.g., Refs \cite{t-1}-\cite{t-6}. 

In addition, very recently, a  program of measurements almost
made to order has been done by the French Atomic Energy Commission. 
They wanted to develop techniques to quantitatively assay low-level
transuranics in bulk solid waste drums.  This was first done by
detecting on-line delayed neutron counting from incident photons 
with energies up to 18 MeV  \cite{fn-1}.   Soon after they did the
same type of analysis, but this time with the delayed neutrons coming
from the simultaneous interrogation of both photons and neutrons
\cite{fn-2}.  (They also interrogated Uranium encased in 
concrete \cite{fn-3}.)


\section{Comments}

The literature presented here, and the literature contained therein, 
will serve as a background to programs like 
\cite{1-1}, which aim to study the detection of fissile materials from
delayed neutrons.  Data on neutron induced fission of nuclei $^{N-1}A$
\cite{n-1}-\cite{n-6} is also of interest as complementary data to the
photofission of nuclei $^NA$. 
 
Finally, all the available information on photofission is an aid 
to  the evaluation of what type
of light source is needed in the near future to best advance the field  
\cite{new}. 

I thank Barry Berman, Mark Chadwick, Dominic Chan, 
Jim Friar, Danas Ridikas, and Bill Wilson for many helpful
comments.  In addition, Fanny Jallu kindly provided references. 
This work was supported by the United States Department of Energy.  



\end{document}